\documentclass{kluwer}    

\newdisplay{guess}{Conjecture}
\usepackage[dvips]{epsfig}

\begin{document}     

\begin{article}
\begin{opening}
\title{Evaporation and condensation processes of giant molecular clouds in
a hot plasma}
\author{Wolfgang \surname{Vieser}\email{vieser@astrophysik.uni-kiel.de}} 
\author{$ $\\ Gerhard \surname{Hensler}\email{hensler@astrophysik.uni-kiel.de}}
\runningauthor{W.\ Vieser \& G.\ Hensler}
\runningtitle{Evaporation and condensation processes of giant molecular 
clouds in a hot plasma}     
\institute{Institut f\"ur Theor.\ Physik und Astrophysik,
Universit\"at Kiel, D - 24098 Kiel, Germany}  
\begin{ao}\\
vieser@astrophysik.uni-kiel.de
\end{ao} 

\begin{abstract}
2D hydrodynamical simulations are performed to examine the
evaporation and condensation processes of giant molecular clouds
in the hot phase of the interstellar medium. The 
evolution of cold and dense clouds ($T = 1000 \mbox{ K}$, 
$n_H = 3 \mbox{ cm}^{-3}$,
$M = 6\cdot10^4 \mbox{ M}_{\odot}$) is calculated in the subsonic
stream of a hot tenuous
plasma ($T = 5\cdot10^6 \mbox{ K}$, 
$n_H = 6\cdot10^{-4}\mbox{ cm}^{-3}$). Our
code includes self-gravity, heating and cooling processes and
heat conduction by electrons. The thermal conductivity of a
fully ionized hydrogen plasma ($\kappa \propto \mbox{T}^{5/2}$) is
applied as well as a saturated heat flux 
in regions where the mean free path of the electrons is large compared to 
the temperature scaleheight. 

Significant differences occur between simulations with and without
heat conduction.
In the simulations without heat conduction, the clouds outermost 
regions is stired up
by Kelvin-Helmholtz (KH) instability after only a few dynamical times. 
This prevents an infiltration of a significant amount of hot gas into the 
cloud before its destruction. In contrast, models
including heat conduction evolve less violently. 
The boundary of the cloud remains nearly
unsusceptible to KH instabilities. In this scenario it is possible to
mix the formerly hot streaming gas very effectively with the cloud material.
\end{abstract}
\keywords{evaporation, condensation, heat conduction, molecular clouds}
\end{opening}  


\section{Introduction}

The Interstellar Medium (ISM) can be described as a clumpy, 
inhomogeneous three-phase medium (\opencite{MO77}: MO77). 
The cold neutral phase ($T \sim 80 \mbox{ K}$, 
$n \sim 40 \mbox{ cm}^{-3}$) is represented by the cores of
molecular clouds which are enveloped by a warm neutral to slightly
ionized medium
($T \sim 8000 \mbox{ K}$, $n \sim 0.3 \mbox{ cm}^{-3}$). These two components
are in pressure equilibrium if the gas is externally heated and can cool
radiatively \cite{F69}. According to MO77 they are embedded in 
a third hot, dilute phase of the
ISM ($T \sim 10^6 \mbox{ K}$, $n \sim 10^{-3} \mbox{ cm}^{-3}$). This
component, which is produced by supernova (SN) explosions, can originally
not be in pressure
equilibrium with the colder phases and has therefore to expand vehemently. 
During this expansion, shocks arise, penetrate through and interact with 
the ambient clumpy
ISM. In OB associations the sequential SNe accumulate and form a simple
giant hot gas bubble, called superbubble, that expands preferentially towards
the halo and occupies a large volume \cite{NI89}. In order to allow for a
more efficient cooling of the superbubble gas the evaporation of incorporated
clouds are invoked \cite{S94}. This evaporation of cloud material 
should be caused by heat conduction of electrons from the hot to the cold
gas. Our aim is to investigate the interaction between the streaming hot
plasma of the ISM and the molecular clouds in order to determine the r\^ole
of heat conduction during the cloud evolution.

A situation described above comes into play in many
astrophysical phenomena. As an example {\it{High-Velocity Clouds}} (HVCs)
are radially falling towards the galactic
disk (see \citeauthor{WW97}, \citeyear{WW97}, for a recent review) and
penetrate through the hot halo gas \cite{K96}. 
Single HVCs are grouped
together in complexes which can be found all over the sky. Interferometer
measurements of the 21 cm {H{\sc i}} line at $1^\prime$\ resolution reveal the
substructure of single complexes that consist of several small clumps
embedded in larger emission regions \cite{WS91}. This, along with
their line widths led Wakker \& Schwarz conclude that the HVCs have
a multi-phase structure consisting of a cold, dense core and a warmer, more
tenuous envelope. Distance measurements for the cloud complexes remain 
difficult.
For at least two of them upper limits for their distance are available
(\citeauthor{D93} \citeyear{D93}; \citeauthor{K95} \citeyear{K95};
\citeauthor{W97} \citeyear{W97}). These estimates reveal that they belong
to the hot galactic halo. 
Although the origin of most of the complexes is still speculative
\cite{B99}, HVCs are
therefore a classic example for structures moving through a hot 
plasma.

Another example for the scenario described above are the much more massive and
larger speculative proto-globular cluster clouds (PCCs) with 
temperatures near $10^4$ K and densities several hundred times that of the 
surrounding gas. Therefore they are gravitationally unstable at masses larger
than $10^6$ M$_{\odot}$.
These clouds originate from condensations of
thermally unstable gas with temperatures of some million Kelvin in the
early epoch of galaxy formation
and can be envisaged
as progenitors of globular clusters \cite{FR85}. The PCCs had to
resist their gravitational collapse for a sufficient time until star formation
ignites. On the other hand, they are accelerated in the gravitational 
potential of the forming protogalaxy and
therefore move through a hot plasma where they become subject to the
growth of Kelvin-Helmholtz (KH) and Rayleigh-Taylor (RT) instabilities
\cite{DR81}. 

Stable models for both, HVCs and PCCs, assuming
hydrostatic and thermal equillibrium consist of large temperature
and density gradients at the surfaces of the clouds where the energy densities
of the hot, tenuous ISM and the warm, dense cloud become equal.
Because of this large temperature gradient and the high temperature of the 
ambient medium of the order of some million Kelvin, heat conduction
has to play an substantial r\^ole on the evolution of such clouds.

\section{Heat conduction}

Heat conduction results in an additional heat flux $\vec{q}$, that can be 
written in the classical case where the mean free path of the electrons is 
small compared to the temperature scaleheight as \cite{S62}
\begin{equation}
\vec{q}_{\mbox{\tiny class}} = - \kappa \cdot \vec{\nabla} T
\end{equation}
with
\begin{equation}
\kappa = \frac{1.84 \cdot 10^{-5} \; T^{5/2}}{\ln \Lambda}
\mbox{ erg s$^{-1}$ K$^{-1}$ cm$^{-1}$}
\end{equation}
where the Coulomb logarithm is
 \begin{equation}
\ln \Lambda = 29.7 + \ln \left [ \frac{T_e}
{\sqrt{n_e} \; 10^6 \mbox{K}} \right ]
\end{equation}
with the electron density $n_e$ and the electron temperature $T_e$,
respectively. The diffusion approximation for the heat flux breaks down
if the mean free path becomes comparable or even greater than the temperature
scaleheight. A common approach is to use a flux limited form, socalled
``saturated'' heat flux. This takes charge conservation into account and yields
results in good agreement with more sophisticated treatments (e.g. 
\opencite{M80}) and with numerical simulations of laser heated plasmas 
(\citeauthor{MN73} \citeyear{MN73}; \citeauthor{MK75} \citeyear{MK75}). 
This  saturated heat
flux takes the form (\opencite{CK77}: CM77)
\begin{equation}
|\vec{q}_{\mbox{\tiny sat}}| = 5 \; \Phi_s \, \rho c^3
\end{equation}
where $c$ is the sound velocity and $\Phi_s$ is a number less than or of 
the order of unity which embodies some
uncertainties concerning the flux limited treatment ($\Phi_s =1$ in our 
calculation). In order to get a smooth
transition between both the classical and the saturated regime we used the
analytical form by \inlinecite{SC92}
\begin{equation}\label{eq3}
\vec{q}=|\vec{q}_{\mbox{\tiny sat}}| \left(
1- \exp \left[ -\frac{\vec{q}_{\mbox{\tiny class}}}
                     {|\vec{q}_{\mbox{\tiny sat}}|}
        \right]
	\right)
\end{equation}
This guarantees that the smaller flux is taken if both differ significantly. 
As a criterion to separate both cases for a cloud of radius $R$ embedded 
in a hot gas with temperature $T_f$, electron density $n_{ef}$ and thus 
the conductivity $\kappa_f$, CM77 introduced
a global saturation parameter which is essentially the ratio of the electron
mean free path to the cloud radius
\begin{equation}
\sigma_0=\frac{0.08 \; \kappa_f \, T_f}{\Phi_s \; \rho_f \, c^3_f \; R}
= \frac{1.23 \cdot 10^4 \; T^2_f}{n_{ef} \; R} \; .
\end{equation}
For $\sigma_0 <
0.027/\Phi_s$ material condenses onto the cloud because radiative losses 
exceed the conductive heat input. 
For $0.027/\Phi_s < \sigma_0 \le 1$ the clouds suffer
classical evaporation, while for $\sigma_0 > 1$ the evaporation is
saturated. The classical evaporation rate is given by CM77:
\begin{equation}\label{eq4}
\dot{m}_{\mbox{\tiny class}} = \frac{16 \,  \pi \; \mu \; \kappa_f \; R}
{25 \;k_B} \;,
\end{equation}
while for the saturated case, $\dot{m}_{\mbox{\tiny class}}$ is multiplied 
by a function $w(\sigma_0)$
\cite{DB93} which lowers the mass loss rate for $\sigma_0 > 0.01$.

\section{Hydrodynamic Simulations}

The evolution of clouds in the subsonic stream of a hot plasma is studied
by two-dimensional hydrodynamical simulations. For this 
the Eulerian equations are solved on a rectangular cylindrical symmetric
``staggered grid''  which is second order in space. 
The boundary conditions are chosen in a manner that material can flow out of 
the computational domain through the upper and right boundary. 
The physical parameters
of the lower boundary that is also the symmetry axis are mirrored. The 
parameters, density and temperature, at the left boundary keep their 
initial values. Also the flow of the plasma is initialized by
setting the velocity of the left boundary to the requested value. 
In order to follow the condensation of the streaming material onto the
cloud a new quantity ``colour'' was introduced. The value of this new
quantity is set in each cell to the density of hot ISM. 
At the beginning of the calculation
only the cells not belonging to the cloud have a colour unequal to zero with
a value equal to its density. During the
calculation this quantity is advected like the others as e.g. density
or energy density. 
For the advection we used the monotonic transport of \inlinecite{L77}. 
The Poisson equation for self-gravity was solved if the new evaluated 
gravitational potential differs from the old one by more than $10^{-4}$. 
The energy equation is extended by heating, cooling and
heat conduction, the latter one being the divergence of the heat 
flux (equ. \ref{eq3}). The adopted cooling function assumed 
collisional 
ionisation equilibrium and
is a combination of the function introduced by \inlinecite{BH89} 
for $T>10^4$K and by \inlinecite{DC72} for the lower temperature 
regime. The heating function considers heating by cosmic rays,
X-rays and the photoelectric effect on dust grains.

\subsection{Theoretical Model}

The initial profile of the cloud is generated for 
hydrostatic and thermal equilibrium. 
After setting the temperature $T_{\mbox{\tiny ISM}}$ and particle density
$n_{\mbox{\tiny ISM}}$ of the hot and tenuous outer medium the energy
density $e_{\mbox{\tiny ISM}}$ of the plasma is known. 
The density and temperature profile
of the cloud is calculated by integrating the equations of hydrostatic
and thermal equilibrium from inside out using the core temperature of 
the cloud as the inner boundary condition and is
truncated when the energy density reaches $e_{\mbox{\tiny ISM}}$.
The model parameters are given in Table~\ref{tab1}. The 
parameters $\sigma_0$ and the evaporation timescale $\tau_{\mbox{\tiny eva}}$ 
are only defined for the simulation with heat conduction.
The hot gas streams with 0.3 Mach.
The resulting $\sigma_0$ for the simulation with heat conduction achieves
a moderate saturation; the evaporation timescale $\tau_{\mbox{\tiny eva}}$ 
is calculated from equation~(\ref{eq4}) correspondingly.
The dynamical timescale as defined to be the sound travel time over
one cloud radius, is $16.2$ Myr. For the grid
$440 \times 160$ cells are used in the
simulation with heat conduction and $640 \times 200$ cells for the
simulation without heat conduction. Thus, the spatial resolution 
amounts to 1.25 pc/cell.
\begin{table}[h]
\caption{Model parameters of the simulation}\label{tab1}
\begin{tabular}{llllllll} \hline
   $T_{\mbox{\tiny ISM}}$  
 & $n_{\mbox{\tiny ISM}}$   
 & $v_{\mbox{\tiny ISM}}$   
 & $R_{\mbox{\tiny cld}}$ 
 & $M_{\mbox{\tiny cld}}$ 
 & density  
 & $\sigma_0$
 & $\tau_{\mbox{\tiny eva}}=M/\dot{m}$
  \\ 
  (K) 
& (cm$^{-3}$) 
& (km s$^{-1}$) 
& (pc) 
& M$_{\odot}$ 
& contrast 
&  
& Myr  \\ \hline  
$5.6 \cdot 10^{6}$ & $6.6 \cdot 10^{-4}$ & $107.4$ & $41.3$ & 
$6.4 \cdot 10^{4}$ & $1.2 \cdot 10^{4}$ & 
$10.5$ & $279$ \\
\hline
\end{tabular}
\end{table}

\section{Results}

\begin{figure}[H]
\centerline{
\epsfig{file=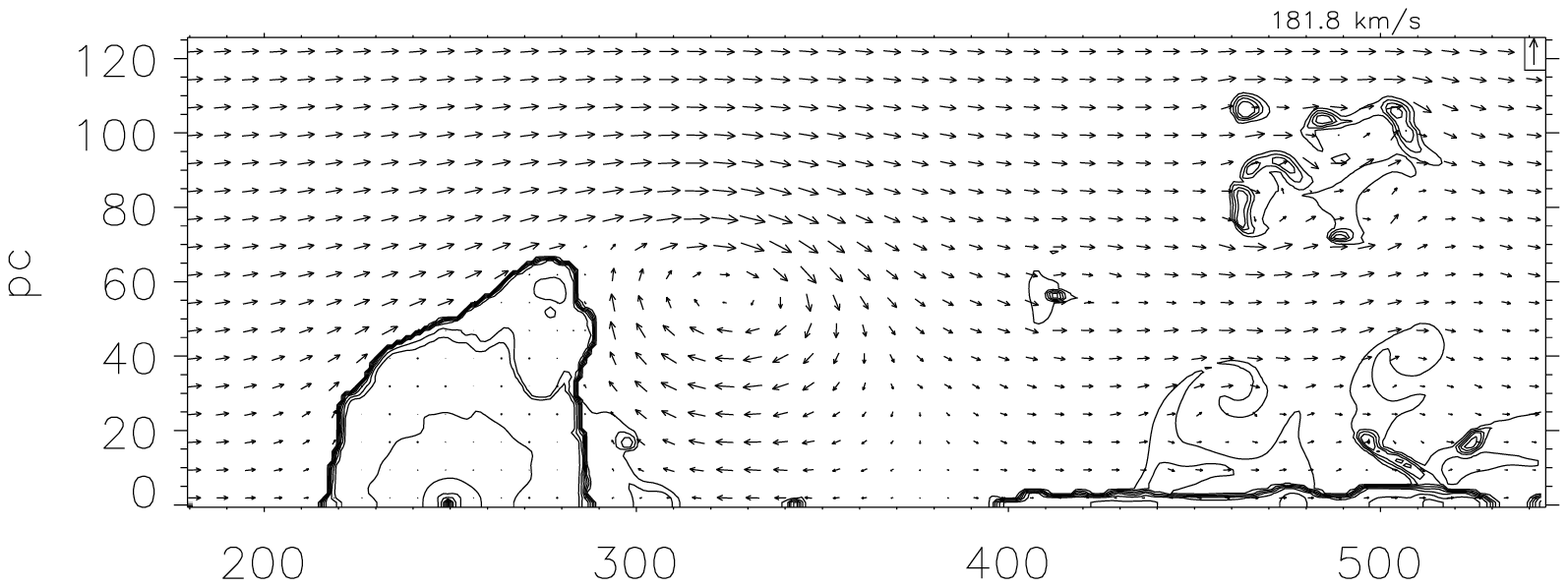,width=27pc}} 
\vspace{-4.0ex}
\centerline{
\epsfig{file=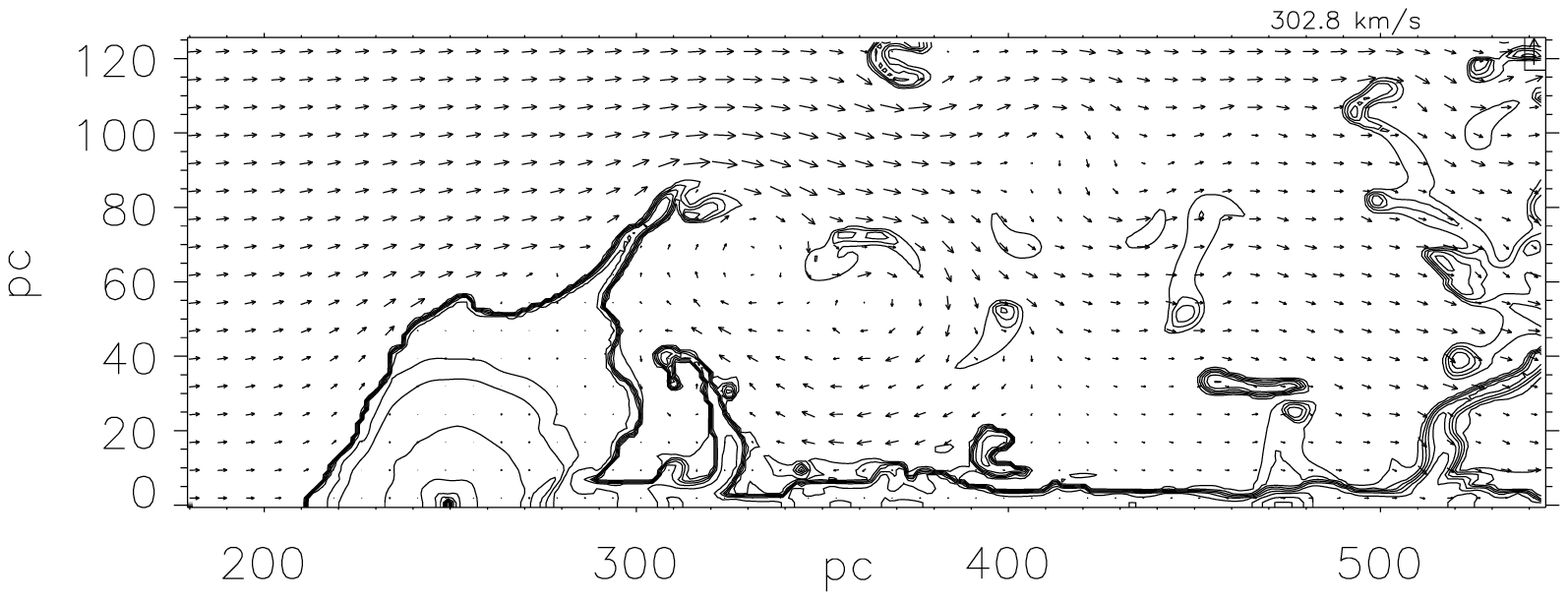,width=27pc}} 
\caption{Density contours for the simulation without heat conduction at
30 Myr (upper panel) and 50 Myr (lower) after the beginning of the 
calculation. The contour lines
represent 5, 10, 50, 100,\ldots $\times \rho_{\mbox{\tiny ISM}}$. The 
velocity arrows scale linearly with respect to the maximum
velocity shown in the upper right.}\label{pic1}  
\end{figure}

\begin{figure}[H]
\centerline{
\epsfig{file=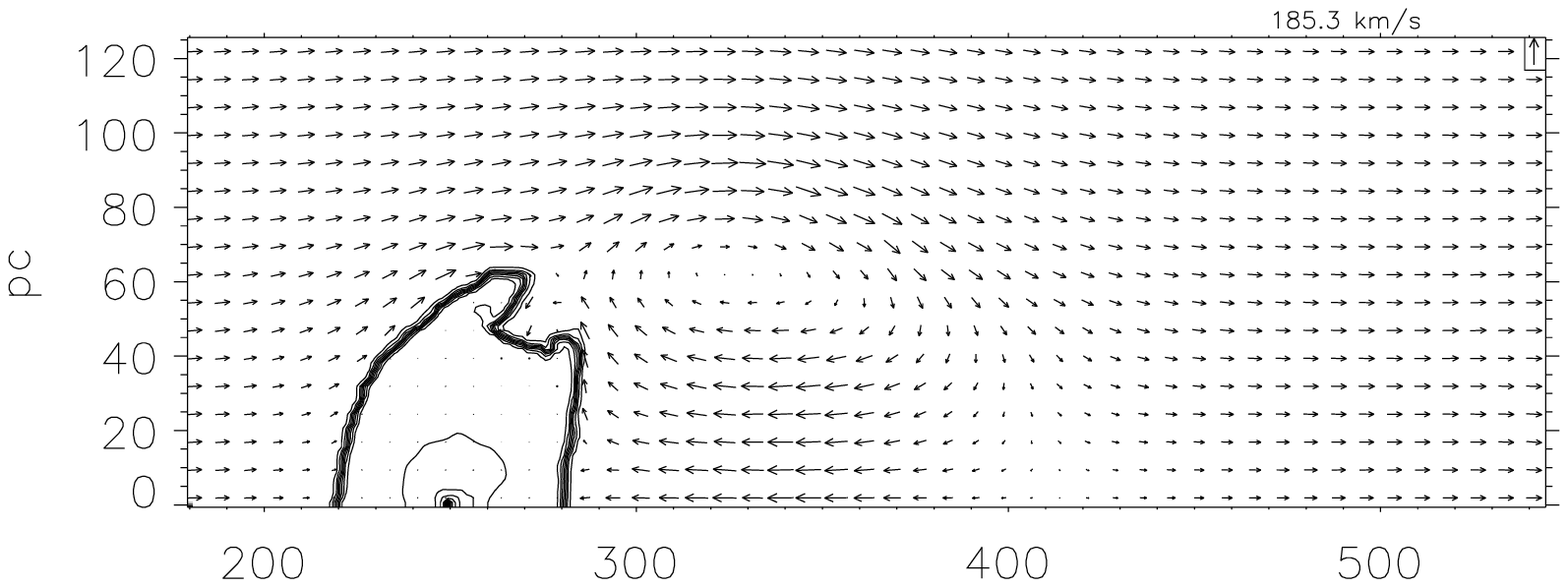,width=27pc}} 
\vspace{-4.0ex}
\centerline{
\epsfig{file=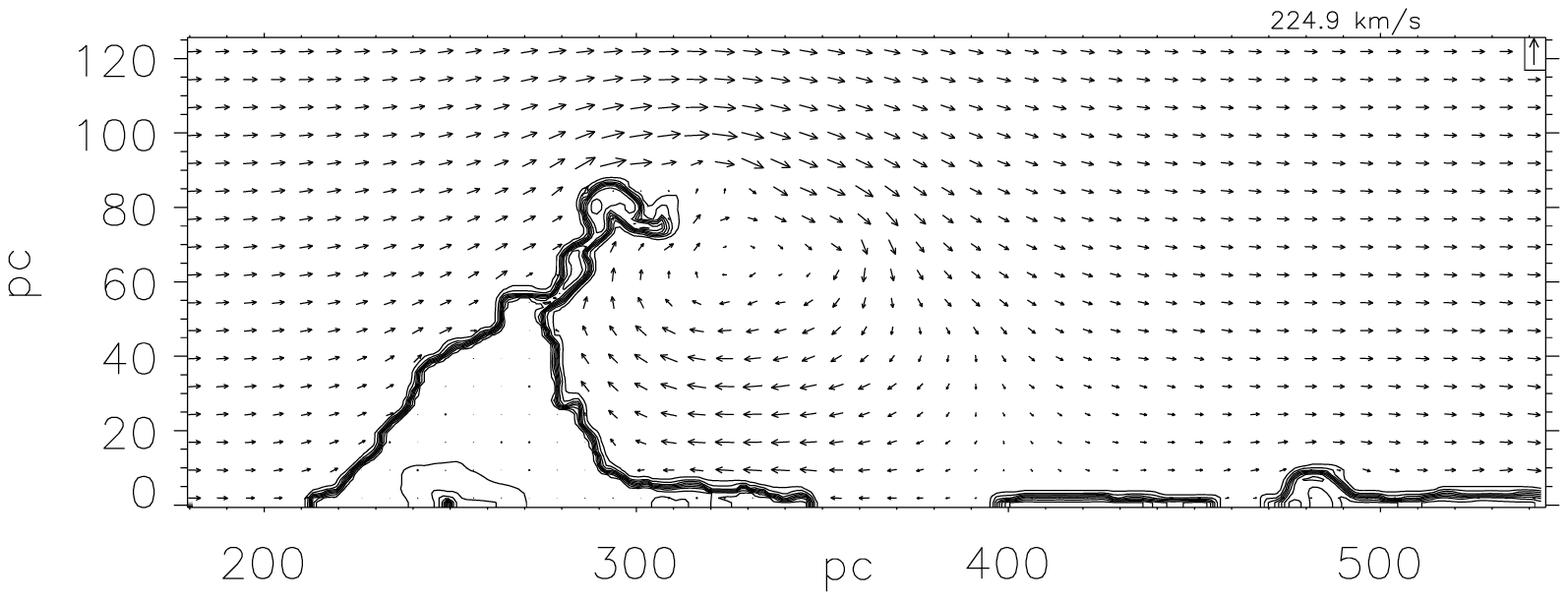,width=27pc}} 
\caption{Density distribution for the simulation with heat conduction at
30 Myr (upper panel) and 50 Myr (lower) after the beginning of the 
calculation.The contour levels are the same as in 
Fig.~\ref{pic1}}\label{pic2}  
\end{figure}

\begin{figure}[H]                  
\centerline{
\epsfig{file=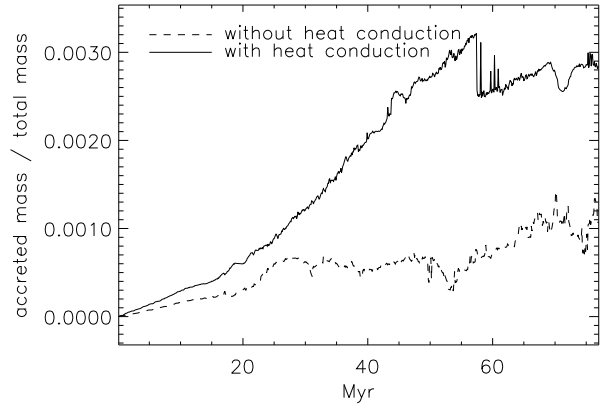,width=20pc}} 
\caption{Evolution of the accretion of the streaming material 
onto the cloud.}   \label{pic3}  
\end{figure}

\begin{figure}[H]                  
\centerline{
\epsfig{file=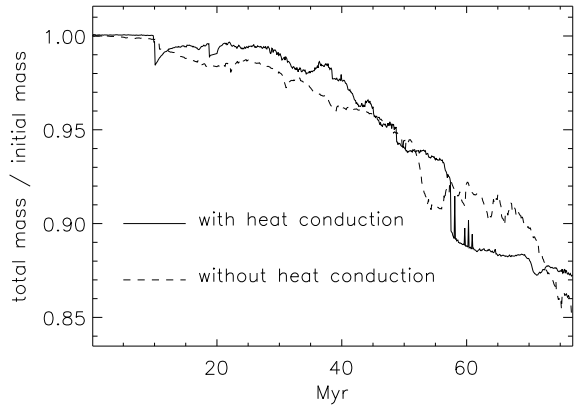,width=20pc}}
\caption{Evolution of the gravitationally bounded mass with the regard to 
the initial mass.} \label{pic4}  
\end{figure}

The inclusion of heat conduction has the tendency to stabilize massive and
large clouds. This is illustrated comparing the density contours at the 
same times (30 Myr and 50 Myr after the beginning of the 
calculations) for the simulations without (Fig.~\ref{pic1}) and with heat
conduction (Fig.~\ref{pic2}).
Without heat conduction the edge of the cloud 
is torn by KH instability already shortly after the beginning when the density
contrast is lowered. Also a very complex velocity structure with many
vortices behind the cloud has formed. On the other hand, heat conduction 
suppresses large scale KH instabilities and only a single large circulation
in the slipstream of the cloud is visible. 
This behaviour can be understood if the conducting electrons move along
the temperature gradient, i.e. radially toward the cloud. This leads to a
deceleration of the contact interphase so that the relative velocity
decreases below a critical value that determines the onset of KH 
instabilities. Also the inner parts of the clouds develop differently.
While in both simulations the cores remain very dense, the model 
without heat conduction develops a compound envelope, with a radially
decreasing density and a diluted outer part of material that is only slightly 
gravitationally bound.
\noindent
The cloud model with heat conduction remains more compact with bound edges. 
In regions where the heat conduction exceeds the cooling the density 
distribution is homogenized while in very dense regions like e.g.\ 
near the core the density structure is unaffected by heat 
conduction. The tendency to smooth out temperature and density inhomogenities
is also the reason why KH instability is suppressed. At the edge of the cloud
an interface builds up which broadens the transition region from the cloud to
the outer medium. This transition region is visible in Fig.~\ref{pic2} as 
a broad band of contour lines where the former steep temperature and
density gradients are diminished. Within this interface at the cloud's 
surface the sheer velocity is therefore lowered and the KH instability 
is reduced, by this, enabling material from the flow to settle 
onto the cloud. This difference is clearly discernible in Fig.~\ref{pic3}. 
After 76 Myr almost 160 M$_{\odot}$ have stuck into the cloud what 
amounts to almost $0.3$\% of the total cloud mass. The
simulation without heat conduction shows an accretion that is only
1/3 of that one with heat conduction because the accreted material is 
stripped off from the cloud again before it can be mixed with deeper 
layers of the cloud. The evolution of the total mass 
(see Fig.~\ref{pic4}) differs only marginally. Long-term simulations 
have to be performed in order to investigate whether the total mass 
reaches a constant value. Nevertheless, it is obvious, that the cloud 
evolution is not suffering evaporation at that stage as claimed by CM77. 
The mass-loss rate corresponding to the analytical expression given
by equ.~\ref{eq4} is by far too high. 
The question has to be addressed whether the assumptions by CM77 are 
valid in such a more realistic case in which instabilities occur and 
a streaming hot plasma serves as an infinite reservoir. 
In order to answer it, a parameter study covering both the condensation
and evaporation regime proposed by CM77 has to be performed and
investigations must include a higher spatial resolution.

\end{article}
\end{document}